# *Ab initio* DFT+U study of He atom incorporation into $UO_2$ crystals

Denis Gryaznov[1*], Eugene Heifets[1], and Eugene Kotomin[2]

[1)] European Commission, Joint Research Centre, Institute for Transuranium Elements, Postfach 2340, Karlsruhe D-76125, Germany

[2)] Institute for Solid State Physics, University of Latvia, Kengaraga 8, LV – 1063 Riga, Latvia

**Abstract**

We present and discuss results of a density functional theory (DFT) study of a perfect $UO_2$ crystals and He atoms in octahedral interstitial positions. We have calculated basic bulk crystal properties and He incorporation energies into the low temperature anti-ferromagnetic $UO_2$ phase using several exchange-correlation functionals within the spin-polarized local density (LDA) and generalized gradient (GGA) approximations. In all these DFT calculations we included the on-site correlation corrections using the Hubbard model (DFT+U approach). We analysed a potential crystalline symmetry reduction and confirmed the presence of the Jahn-Teller effect in a perfect $UO_2$. We discuss also the problem of a conducting electronic state arising when He is placed into a tetragonal antiferromagnetic phase of $UO_2$. Consequently, we found a specific lattice distortion which allows us to restore the semiconducting state and properly estimate He incorporation energies. Unlike the bulk properties, the He incorporation energy strongly depends on several factors, including the supercell size, the use of spin polarization, the exchange-correlation functionals and on-site correlation corrections. We compare our results for the He incorporation with the previous shell model and *ab initio* DFT calculations.

PACS: 71.15.Ap, 71.15.Mb, 71.15.Nc, 71.20.-b, 71.27.+a, 71.70.Ej

[*] Corresponding author: Tel.: +49-7247-951462, *E-mail address:* gryaznov@mail.com (D. Gryaznov)

# 1. Introduction

$UO_2$ is one of the main materials for nuclear reactor fuel. A significant amount of defects are formed in the fuel under operational and storage conditions, which affects fuel properties and performance. In particular, He atoms are produced, mainly as a result of α-decay. The concentration of He can be as high as 1%, which, depending on the temperature and fluence, can contribute to gas bubble formation [1], swelling and gaseous release, which might lead to fuel rod failure. Despite its obvious technological importance, there is still discussion about the magnitude of the calculated He incorporation energy into the $UO_2$ matrix.

The first successful atomistic modelling with an estimation of the incorporation energy of He in $UO_2$ was carried out in the early 1990s [2], using semi-empirical interatomic potentials and the shell model. Corresponding *ab initio* calculations were started a decade later (see [3] and references therein). It should be stressed that first studies on perfect and defective $UO_2$ based on *ab initio* methods mostly neglected the electron correlation correction effects [4-7], which leads to a wrong conclusion that this is a metal instead of a semiconductor. Moreover, even spin polarization was not always taken into account, which is in contradiction with the experimental fact that $UO_2$ is an anti-ferromagnetic (AFM) material at temperatures below the Neel temperature $T_N \sim 30.8$ K [8]. The shell model estimate of the *incorporation energy* of He in $UO_2$ [2] (–0.1 eV, i.e. the incorporation is energetically favourable) was exactly reproduced recently in the DFT pseudopotential-plane wave study [6] with the ABINIT code neglecting the correlation corrections. We will discuss this (fortuitous) agreement below.

In the present study, we try to re-consider these estimates, discuss the role of structure distortion in defect studies and check how the results depend on the choice of the supercell, the exchange-correlation functional, and on-site electron correlation corrections. This became possible thanks to the development of theoretical methods allowing a proper treatment of strongly correlated systems, such as actinides, in general, and $UO_2$, in particular. The on-site correlation corrections can be included in the calculations using either the DFT with an energy correlation correction (DFT+U), e.g. in the formalism by Dudarev et al. [3]), or via so-called DFT-HF *hybrid* functionals. The latter approach provides very good results for the band gap of bulk $UO_2$ [9] which in the standard DFT turns out to be a metal. However, both methods have considerable shortcomings. The DFT+U accounts for the correlations on a specific orbital only, whereas the hybrid functionals face difficulties in implementation in the plane wave computer codes and require significant additional computational expenses.

This is why hybrid functionals are available currently for *f*-elements only in the LCAO (linear combination of atomic orbitals) basis set codes, such as Gaussian2003 [10] and CRYSTAL06 [11-12]. In its turn, the DFT+U approach is now available in most plane wave codes and has already been applied for actinides [13-14]. In particular, the *ab initio* total-energy and molecular-dynamics program VASP [15-17] also demonstrates very good results in perfect $UO_2$ studies [18-19]. So far, the DFT+U approach was applied for native point defects in $UO_2$ [20] but not for He impurities.

The present paper is organized as follows. In Section 2 we introduce the computational details. Sections 3.1 and 3.2 are devoted to our main results for the $UO_2$ bulk and He incorporation, respectively. Section 3.2 is divided into two parts, namely He incorporation is analyzed in non-distorted tetragonal and distorted $UO_2$, respectively. Conclusions are presented in Section 4.

## 2. Method

In our study we used the VASP computer code [15-17, 21] and compared several exchange-correlation functionals: LDA [22] and two GGA functionals, namely PW91 [23] and PBE [24]. Such a comparison is important for reliable calculations of the defect energetics [25]. The computations were performed with spin-polarized techniques, except for the cases where we tested properties for non-magnetic electronic configurations. In the present computations, we used the DFT+U approach [3, 26]. It provides good ground-state properties of $UO_2$ and other strongly correlated systems, e.g. transition metal oxides, including the band gap and magnetic properties. In particular, a simplified version of this approach proposed by Dudarev has already been successfully applied to $UO_2$ [3, 18-21] and other $f$-electron compounds such as $CeO_2$ [27]. An advantage of this DFT+U version is that the correlation (U) and exchange (J) parameters do not enter separately the Hamiltonian; only the difference U-J is essential. The value of the correlation energy (U) was fixed at 4.6 eV in our calculations, as suggested in Ref. [28] (based on the X-ray photoemission spectroscopy combined with Bremsstrahlung isochromat spectroscopy measurements). The value of the exchange parameter (J) was fixed at 0.5 eV as proposed by Dudarev [3].

We used the projector augmented wave (PAW) method [29] and scalar-relativistic effective core potentials (ECPs) [30] as implemented in the VASP code substituting for 78 core electrons on U atoms and 2 core electrons on O atoms. A recent comparison of the full potential and frozen-core approximations demonstrated that the PAW ECPs give very reasonable results [31]. The plane wave cut-off energy was fixed at 520 eV, except for defective supercell calculations with the PBE functional, where it was increased up to 621 eV. As was mentioned above, at low temperatures $UO_2$ has the AFM structure. We do not employ any vector description for spins, neglect also spin-orbit interaction as well as a dynamic Jahn-Teller effect as discussed in the literature [32]. Instead we use a simplified model assuming that the U spins are parallel within the (001) planes and alter their signs in the [001] direction. The corresponding crystalline structure is characterized by the tetragonal unit cell, consisting of two fluorite primitive unit cells. Similar approximation was used in recent theoretical studies [33]. This corresponds to the so-called *1-k* magnetic structure. The only published attempt of a first-principle study for a more complex *3-k* magnetic order in UO$_2$, which was suggested experimentally [32], was performed in Ref. [34]. However, in the latter paper [34] the energy differences between *1-k* and *3-k* phases appeared to be very tiny and the relative stability of these phases depends on technical details (a variant of so-called double counting correction) of the DFT+U method. Because of uncertainties of these results, we decided to leave more complex non-collinear *3-k* magnetic structure outside of scope for our present study and perform our simulations using much simpler *1-k* structure.

As it is described below, we found that it was important to investigate a possibility of a UO$_2$ crystal symmetry reduction. To this end, we re-optimized lattice constants using the same unit cell but lifted the symmetry constrain. Then we set to zero those lattice constant components which had a very small value (< 0.0005 Å) and re-optimized again with imposed symmetry constrains. Low-symmetry simulations were performed with the PBE functional only. Integrations over the Brillouin zone (BZ) employed 6x6x6 Monkhorst-Pack [35] k-point mesh in this case, and the electron occupancies were determined with the Gaussian method using the smearing parameter of 0.25 eV.

The bulk properties of UO$_2$ were calculated using the non-distorted tetragonal unit cell only. For high precision of computations, we applied the tetrahedron technique [36] when calculating the electron states occupations, and integration over the BZ was performed using 9x9x7 *k*-set mesh centered at the Γ-point of the BZ. The bulk modulus was estimated by fitting parameters of the Birch-Murnaghan equation of state [37] to the adiabatic energy profiles (the total energy per unit cell *vs.* unit cell volume) calculated with the VASP code. The bulk properties were calculated for optimized lattice parameters in the AFM state.

In simulations of He atom incorporation into a $UO_2$ crystal, we consider only the octahedral interstitial position which is believed to be the energetically most favourable position. Recently, Garrido et al. [38] confirmed such a preference experimentally on the basis of the channelling Rutherford backscattering spectroscopy technique combined with nuclear reaction analysis. In this interstitial position, He atoms are situated in the center of oxygen cubes of the fluorite lattice (Fig. 1). We used two supercells of different sizes for modelling He incorporated into $UO_2$. The small supercell (Fig. 1a) is a fluorite crystallographic unit cell, containing 4 $UO_2$ molecules and a He atom. The AFM ordering reduces the symmetry of this supercell from cubic to tetragonal. A larger supercell consists of two fluorite crystallographic cells with a single He atom incorporated into one of the interstitials (Fig. 1b) and thus contains 25 atoms. These two supercells correspond to 25% (13-atom supercell) and 12.5% (25-atom supercell) effective He concentrations. Note, the calculations without any symmetry by using even larger supercells would require unrealistic computer time. The integrations in the reciprocal space over the BZ were performed using a 9x9x9 mesh of the *k* points for 13-atom supercell and a 9x9x7 mesh of *k* points for 25-atom supercell and the tetrahedron method for electronic occupancies. Both the *k*-meshes were centered at the Γ-point of the BZ. Along with four functionals used for the present simulations, we performed also test calculations for a combination of the ECPs generated by means of the LDA functional with the energy computed employing the PBE functional. These tests were performed in order to better understand the results of Ref. [6]. To find the correct low-symmetry crystal structure for $UO_2$ with incorporated He atoms, we applied to each supercell the same procedure as described above for a pure $UO_2$. These simulations were done using the PBE functional. All calculations of density of states (DOS) in this work required re-calculation of the electronic structure using the tetrahedron method with given occupations of the electron states.

## 3. Results and discussion

### *3.1. Bulk properties*

Our DFT+U computations predict that the energy of the $UO_2$ crystal in the AFM state is lower than in the ferromagnetic (FM) state, in agreement with experimental data. As it was already mentioned, the AFM ordering reduces the cubic symmetry of the $UO_2$ fluorite structure to tetragonal. In our study, the tetragonal unit cell was fully relaxed, in order to calculate the bulk properties. As a result, the optimized lattice constant *c* (Table 1) along the z-direction with alternating U spins is reduced by ~1% as compared to the two other directions *a*, i.e. the structure remains very close to a cubic one, in agreement with experimental data [39]. Overall, the lattice cubic constants obtained with the LDA functional are slightly smaller than the experimental constant of 5.47 Å [40]. However, the GGA calculations overestimate it, following a well known general trend.

The calculated cohesive energy $E_{coh}$ (Table 1) is very close for the PW91 and PBE functionals, being slightly smaller than that for the LDA. This is not surprising since the LDA is known to overestimate $E_{coh}$. Note that the previous theoretical study by Dudarev et al. [3] based on the LDA+U approach, combined with a linear muffin tin orbital method of the electronic structure computations, gave even better agreement with the experiment (Table 1).

Values of the bulk modulus B are smaller for the two GGA functionals in comparison with that for the LDA functional, which is again in line with common trends in the DFT calculations. The LDA functional provides the B value close to experiment but again differs significantly from previous theoretical studies. In particular, Dudarev et al. [3] obtained a considerably smaller B value, whereas non-spin-polarized LDA calculations [41] suggested 252 GPa which significantly overestimates the experimental value. Notice that all theoretical values presented in Table 1 for a comparison with our results were taken from spin-polarized calculations only.

The effective atomic charges $Q_{eff}$ in $UO_2$ have not been discussed so far in the literature. Our effective U ion charges in Table 1 were calculated using the topological (Bader) method [42-43]. These charges considerably differ from the formal charges ($U^{4+}$, $O^{2-}$), thus, suggesting a partly covalent nature of $UO_2$ bonding. The results for all functionals applied in this study agree very well with the effective charge of U and the U-O bond covalency. The magnetic moments μ on the U ion confirm this picture.

The DFT+U approach gives the $UO_2$ band gap $E_g \sim 2$ eV, consistent with the experiment (Table 1). Both the GGA functionals (PW91 and PBE) suggest the same values, whereas the LDA functional gives a slightly smaller gap. The band gap of 1.3 eV obtained by Dudarev et al. [3] is considerably underestimated. At any rate, this is in contrast to the zero gap obtained with the standard DFT neglecting the on-site electron correlation corrections.

The total density of states (DOS) together with the partial DOS is presented in Fig. 2. This confirms the $UO_2$ semiconducting nature. As it is expected, the U *5f* electrons form a separate (localized) band below the Fermi energy with a small contribution of O *2p* orbitals, which leads to the U-O partial bond covalency discussed above. This was observed also in the LCAO computations with the Gaussian code [9, 33] performed using the hybrid functional. However, the sub-gap between the U *5f* band and O *2p* band (~0.2 eV) in our calculations is approximately an order of magnitude smaller than the value obtained with the hybrid functionals, which is in better agreement with the experiment [28]. The results presented here clearly demonstrate that the plane wave approach within the PAW ECPs provided in the VASP code leads to very reasonable results and can, thus, be applied to defect and He calculations in $UO_2$.

Allen [44] and Caciuffo et al. [32] demonstrated the existence of the Jahn-Teller effect in a pure $UO_2$ at temperatures below 200 K. Therefore, we tested theoretically a reduction in the crystalline symmetry due to a crystal deformation as a logical step in proper theoretical studies. The structure with the lowest energy appears to be orthorhombic with the point group symmetry $D_{2h}$. This structure is formed from the tetragonal one by compressing along the vector parallel with the magnetic moments on uranium ions, i.e. along the z-direction in our case ($c$ = 5.508 Å) whereas two other lattice constants become slightly different ($a$ = 5.575 Å and $b$ = 5.576 Å). Such a tiny difference is enough to identify the symmetry properly. The structure was characterized by the same bulk properties and did not show any significant difference with the fluorite one used in the beginning of our study (Table 1). However, the total energy gain for the reduced symmetry structure is of the order of ~0.01 eV, which demonstrates that distortions in tetragonal $UO_2$ and the presence of the Jahn-Teller effect may not be ignored.

*3.2. He incorporation*

*3.2.1. Tetragonal structure*

He atom incorporation into the octahedral interstitial position of the tetragonal lattice of $UO_2$ crystal does not lead to a significant lattice relaxation, regardless of the functional used. The largest displacements were obtained for U atoms located in the direction of the spin alternation and even these are smaller than 0.5% of the lattice constant (Table 2). The nearest O and U atoms relax outward from the He atom. The lattice constant $c$ is increased due to the He presence by ~0.04 Å for the smaller 13 atom supercell and by ~0.02 Å for the larger 25 atom supercell, whereas the lattice constants $a$ in the basal plane contract by ~0.01 Å in both supercells.

The He *incorporation energies* were calculated according to the definition proposed by Grimes and Catlow [45] as the total energy of a supercell with inserted He atom minus a sum of the energy of the same perfect supercell and that for an isolated He atom. This requires the calculation of non-defective $UO_2$ with the lattice parameter optimization and defective $UO_2$ with the full lattice relaxation for the corresponding supercell. The incorporation energies for different functionals are summarised in Table 3. The calculation of the incorporation energy for the non-magnetic $UO_2$ (denoted as NM) for the 25 atom supercell was also performed for a comparison with previous results [5-6].

Within the GGA+U approach (PW91 and PBE functionals) and PAW-ECP the incorporation energy considerably decreases, from 1.6 eV in 13 atom supercells down to 0.8 eV in 25 atom supercells. There is good agreement between the GGA functionals used. That is, the incorporation energy for He in $UO_2$ remains positive in these calculations. This significantly differs from the results of the classical shell model [2] and of previous GGA calculations for the non-magnetic $UO_2$ [6] which surprisingly predict the same negative incorporation energy (-0.1 eV). However, this agreement seems to be an artefact. Indeed, we found that it is the combination of PAW-ECPs generated with the LDA functional and simulations employing the PBE functional for the total energy which leads to the reported incorporation energy. Moreover, even more negative incorporation energy could be obtained if the supercell size increases. However, computations of the non-magnetic state with a consistent choice of PAW ECPs generated with PW91 functionals and the *same* functional used for the total energy calculations, lead to a quite different, *positive* He incorporation energy (1.3 eV for the 25 atom supercell). This coincides with the results [5], but is significantly larger than the incorporation energy obtained in the present simulations for the AFM state with the correlation corrections included. This demonstrates that an appropriate account of spin-polarization is necessary for obtaining correct results.

More detailed analysis of the electron structure of calculated supercells, however, provided a surprise. We found that the band containing U *5f* electrons is located *directly* at the Fermi level. Such a conducting state is surely unexpected for He atoms with strongly bound *1s* electrons. In other words, the obtained electronic structure of the supercell with incorporated He atom thus turned out to be wrong. Consequently, further investigations of the distortion mechanisms for defective $UO_2$ were necessary.

### *3.2.2. A distorted structure*

A solution of the above-mentioned problem was found through lifting the system symmetry by means of a small distortion. We performed the supercell geometry optimization without imposed symmetry. As a result, the total energies of the supercells decreased and the electronic state became semiconducting. The calculation, as previously, was spin-polarized with the AFM alignment of spins on U ions. Since these computations are very time consuming, we limited ourselves to the PBE exchange-correlation functional only. The obtained crystal structure appeared to have a *monoclinic* symmetry with the He atom occupying the interstitial position with $C_{2v}$ point symmetry. This structure can be described as the tetragonal supercells sheared along base diagonal (as shown by the arrow in Fig. 1a). Still, the deviation of the lattice from a cubic structure appears to be very small. The calculated lattice vectors for the 13 atom supercell in the basal plane are $a=b = 5.564$ Å (with the angle between them $\gamma = 90.08°$) whereas $c = 5.563$ Å (with the angle with each of the basal lattice vectors is $\alpha = \beta = 90.17°$).

The DOS for 13 atom *monoclinic* supercell is shown in Fig. 3. The total DOS for the same supercell of a pure $UO_2$ (without He atom) is also added there for a comparison. The band gap in the supercells with He atom is ~2.5 eV, which is ~0.5 eV higher than that for pure $UO_2$. He *1s*-states lie ~7 eV below O *2p*-states, which is much lower than valence electrons. Like in pure $UO_2$, the U *5f* electrons form both the highest valence band and the lowest conduction band. The introduction of a He atom into an interstitial position leads to the higher-energy shift by ~0.5 eV of both the O *2p* valence band and the conduction band.

To estimate the He incorporation energy, we had to use supercells for a perfect $UO_2$ crystal with the same reduced (orthorhombic) symmetry which was found at the previous stage of our study as described in Section 3.1. The re-calculated incorporation energies turn out to be now 0.59 eV and 0.18 eV for 13 and 25 atom supercells, respectively. Both values are much smaller than those obtained using the tetragonal supercells (Table 3); for the 25 atom supercell the incorporation energy is smaller than that for the 13 atom supercell. This suggests repulsion of He atoms incorporated in the neighbouring interstitial positions. However, the obtained incorporation energies are still noticeably larger than the value estimated using the shell model [2]. We expect that the larger supercells could give in principle slightly smaller values for the He incorporation energy, but such computations are currently impossible for us with available computational resources. The more so, there is another important factor: shell model simulations are capable to account for van der Waals (dispersion) interactions, unlike the DFT calculations. In particular, it was shown recently by Mourik et al. [46], that the current DFT methods do not describe properly the rare-gas dimers. Therefore, we assume that our values of the He incorporation energy may be somewhat overestimated. As far as we know, DFT-based estimates of van der Waals interactions in solids are currently under development (see, for example, [47]). Taking into account *a posteriori* shell model estimate of van der Waals contribution to the He incorporation energy of ~ -0.25 eV [48], the corrected value for the 25 atom supercell becomes ~ -0.07 eV, which indeed is very close to -0.1 eV obtained in the calculations with the classical interatomic potentials [2, 49]. This agreement demonstrates a potential convergence of incorporation energy with respect to the supercell size. The negative value of the He incorporation energy indicates the preference of He to remain inside $UO_2$ at low temperatures.

The obtained reduction of the supercell symmetry for a He atom embedded into $UO_2$ clearly indicates presence of the Jahn-Teller effect also in this case. Since the He atom is a spherical closed-shell system with a very strongly bound pair of electrons, it cannot induce itself the Jahn-Teller instability. Therefore, the introduction of He atoms just facilitates the existing instability in a pure crystal and modifies the symmetry of the stable crystal structures.

**Conclusions**

We presented and discussed here the results of DFT+U simulations of a pure $UO_2$ in the low-temperature AFM phase and He atom incorporation into the octahedral interstitial position therein. We found that a pure crystal reveals small orthorhombic distortion and the Jahn-Teller effect, in agreement with experimental data [32, 44] and recent hybrid functional LCAO calculations [51]. Introducing He atom into the tetragonal supercells of the $UO_2$ crystal leads to a conducting solution to the Kohn-Sham equations, whereas a monoclinic distortion allows for restoring the correct semiconducting state. Appearance of this distortion also suggests the Jahn-Teller effect for the He impurity and further confirms the importance of the lattice distortion for defect incorporation studies in $UO_2$ (and probably, other actinide). Our estimate shows the energy gain while incorporating He atoms into uranium dioxide.

We have also shown that spin polarization cannot be neglected in $UO_2$ simulations. Finally, mixing in *ab initio* computations of the GGA functional with ECPs obtained with the LDA functional should be avoided. Such a mixture gives incorrect results substantially different from the simulations which use the same functional for both ECPs and electronic structure computations. As a futher step, we plan to consider the effect of spin-orbit interaction and *3-k* magnetic structure.

**Acknowledgements**

This study was supported by the EC Framework Programme 7 F-BRIDGE project, the ACTINET JP-07-02 project, and the Proposal Nr. 25592 from the EMS Laboratory of the PNNL. Authors are greatly indebted to R.W. Grimes, R. Caciuffo, R.A. Evarestov, R. Konings, T. Wiss, E. Maugeri, D. Sedmidubsky and M. Freyss for very useful discussions. D.G. acknowledges also the EC for support in the frame of the Program "Training and mobility of researchers". Many thanks to P. Van Uffelen for our invitation to ITU, Germany, introduction into the problems of nuclear fuel modelling and a strong support which made our work successful.

TABLE 1. A comparison of UO$_2$ bulk properties (lattice constants (*a, b, c*), cohesive energy E$_{coh}$, bulk modulus B, uranium effective charges Q$_{eff}$(U) and magnetic moments μ(U), and band gap E$_g$) obtained with different DFT exchange-correlation functionals E$_{xc}$ (see the text for details).

|  | Tetragonal phase | | | Orthorhombic phase | Expt | Other calculations |
| --- | --- | --- | --- | --- | --- | --- |
|  | LDA+U | PW91+U | PBE+U | PBE+U | | |
| a (b) Å | 5.463 | 5.562 | 5.567 | 5.576 (5.575) | 5.47[c] | 5.55[f], 5.37[a], 5.45[b], 5.46[b] |
| c, Å | 5.418 | 5.511 | 5.512 | 5.508 | | 5.47[f] |
| E$_{coh}$, eV | 26.0 | 23.1 | 23.0 | 23.0 | 22.3[a] | 22.2[a] |
| B, GPa | 196 | 183 | 180 | - | 207[e] | 173[a], 219[b] |
| Q$_{eff}$(U), e | 2.6 | 2.6 | 2.6 | 2.7 | | |
| μ(U), μ$_B$ | 1.7 | 2.0 | 2.0 | 2.0 | 1.74[c] | ~1.9[a] |
| E$_g$, eV | 1.75 | 1.94 | 1.94 | 2.0 | 2.0[d] | 1.3[a], 3.13[b], 2.64[b] |

[a] Ref. [3]
[b] Ref. [39]
[c] Ref. [32]
[d] Ref. [28]
[e] Ref. [50]
[f] Ref. [20]

TABLE 2. Displacements of atoms in the first two spheres around a He atom incorporated into an octahedral interstitial. Only one representative atom is given from each set of symmetrically equivalent atoms. The origin of coordinates is set at the He atom. Original positions of atoms are given as fractions of lattice constants and the displacements are expressed in percentages of the same lattice parameters *a* and *c* of the conventional unit cell (see Fig. 1 ).

| atom | x, *a* | y, *a* | z, *c* | 13 atom supercell | | | 25 atom supercell | | |
|---|---|---|---|---|---|---|---|---|---|
| | | | | Δx, % *a* | Δy, % *a* | Δz, % *c* | Δx, % *a* | Δy, % *a* | Δz, % *c* |
| U1 | 0.0 | 0.0 | 0.5 | 0 | 0 | 0 | 0 | 0 | 0.43 |
| U2 | 0.5 | 0.0 | 0.0 | 0 | 0 | 0 | 0 | 0 | 0 |
| U3 | 0.5 | 0.5 | 0.5 | 0 | 0 | 0 | 0 | 0 | -0.12 |
| O1 | 0.25 | 0.25 | 0.25 | 0.12 | 0.12 | 0.07 | 0.16 | 0.16 | 0.05 |
| O2 | 0.25 | 0.25 | 0.75 | - | - | - | 0.01 | 0.01 | 0.26 |

TABLE 3. The incorporation energies (eV) calculated for different combinations of ECPs and exchange-correlation functionals ($E_{XC}$). Calculations for the AFM state include the on-site correlation corrections.

| PS | $E_{XC}$ | 13 atoms | 25 atoms | |
|---|---|---|---|---|
| | | AFM | NM | AFM |
| LDA | LDA | 1.83 | - | 1.24 |
| PW91 | PW91 | 1.64 | 1.33 | 0.77 |
| PBE | PBE | 1.57 | 1.35 | 0.81 |
| LDA | PBE | -0.11 | -0.14 | -0.85 |

Figure captions.

FIG. 1. a) A 12 atom unit cell of UO$_2$ with He atom inserted into an interstitial octahedral position. The arrow above the unit cell shows the direction of monoclinic deformation obtained in our simulations (see Section 3.2.2). b) A 25 atom unit cell with a He atom in one of interstitial positions.

FIG. 2. The total and partial DOS calculated using the PBE functional for ideal UO$_2$. The Fermi energy is taken as zero. The negative values of densities of states correspond to spin down electrons.

FIG. 3. DOS obtained for 13 atom AFM supercell with two reduced symmetry structures obtained in the present study (see text for explanations). The Fermi energy is taken as zero. The negative density corresponds to spin down electrons. Only the largest contributions to each band are indicated.

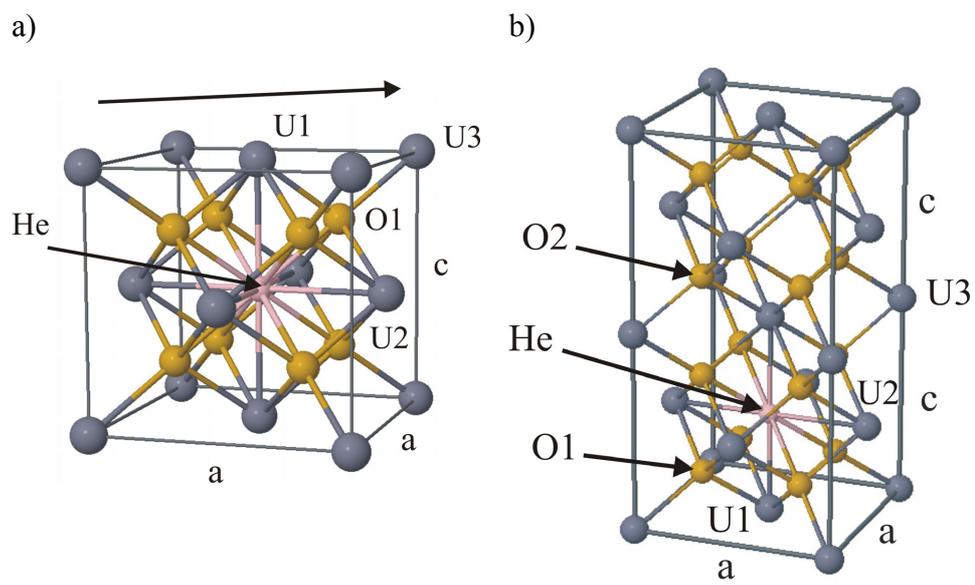

FIG. 1

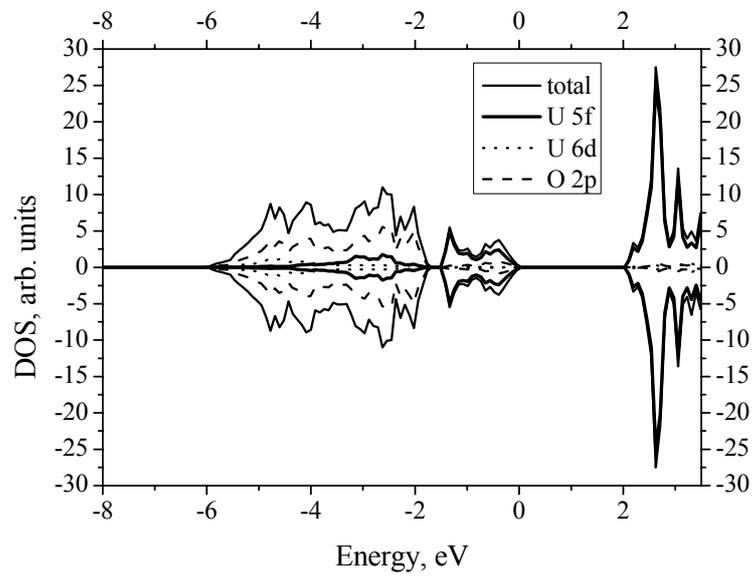

FIG. 2

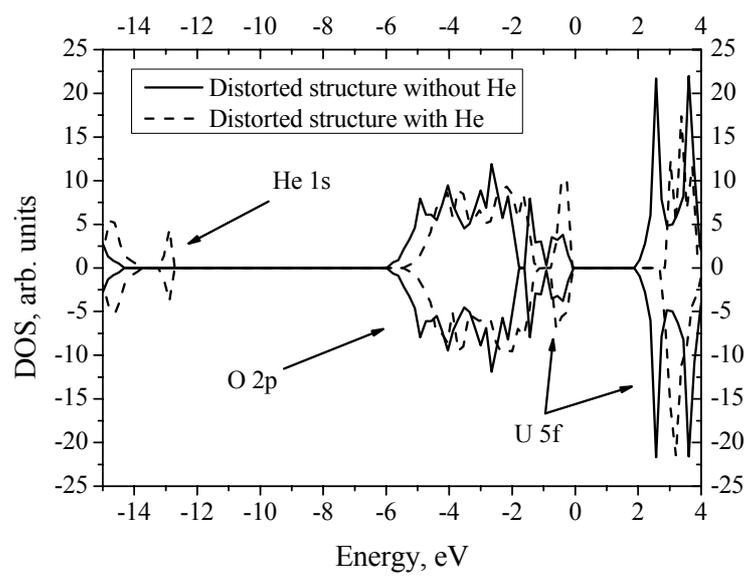

FIG. 3